\documentclass{iau}

\usepackage{graphicx}
\usepackage{alltt}
\usepackage{wasysym} 
\usepackage{amssymb}

\begin{document}
	
	
\title[Tides and Exoplanets]{On Tides and Exoplanets}

\author[S.Ferraz-Mello]   
{S.Ferraz-Mello}

\affiliation{Instituto de Astronomia, Geof\'{\i}sica e Ci\^encias Atmosf\'ericas, Universidade de S\~ao Paulo, Brasil \\ email: {\tt sylvio @ usp.br}}

\pubyear{2022}
\setcounter{page}{1}
\jname{Multi-scale (time and mass) dynamics of space objects} 
\editors{A. Celletti, C. Beaug\'e, C. Gale\c{s}, A. Lemaitre, eds.}
\volume{364}

\maketitle
\def\hig{}
\begin{abstract}
	This paper reviews the basic equations used in the study of the tidal variations
of the rotational and orbital elements of a system formed by one star
and one close-in planet as given by the creep tide theory 
and Darwin's constant time lag (CTL) theory. At the end, it reviews and discusses the determinations of 
the relaxation factors (and time lags) in the case of host stars and hot Jupiters based on actual observations of orbital decay, stellar rotation and age, etc. It also includes a recollection of the basic facts concerning the variations of the rotation of host stars due to the leakage of angular momentum associated with stellar winds.

\end{abstract}

\section{Introduction}

Our current knowledge of dynamical tides is mainly based on Darwin's theory of 1880. All studies done before were based on the static or stationary ellipsoidal models of Jacobi, Maclaurin and Roche and were focused on the relative equilibrium of the forces acting on the considered bodies. 

Darwin was the first to take into account that one body's response to tidal forces is not instantaneous, but suffers a delay that depends on the viscosity of the body.

His first approach using 
hydrodynamical equations, Darwin (1879) showed that fluid bodies respond to tidal torques with a lag that, in case of viscous bodies with small viscosity, is proportional to the frequency of the torque.
In his 1880 theory, Darwin (1880) used this result as an insight to assume that the Earth's tides, due to the stresses generated by the lunar attraction, have an 'ad hoc' lag proportional to the tide frequency and calculated the secular changes in the orbital elements of the Moon. 
This approach is followed even today. It is the basis of the so-called CTL (constant time lag) theories.

In the extended reformulation of Darwin's general theory by Kaula (1964), the 'ad hoc' lags were kept arbitrary, thus allowing for variants of Darwin's theory, as the CPL (constant phase lag) theory (MacDonald, 1964; Jackson et al. 2008).
It also opened the way for considering more complex laws regulating the lags and
allowed the construction of theories able to describe dissipation in rocky planets 
(Efroimsky and Lainey, 2007; Gevorgyan et al., 2020). 

An alternative model valid for both gaseous and stiff bodies, the creep tide theory, was proposed by Ferraz-Mello (2012, 2013). It is based on an approximate solution of the Navier-Stokes equation and is equivalent to Darwin's theory in the case of viscous bodies. In this theory, the actual surface of the body creeps towards the instantaneous equilibrium surface with a speed proportional, at each point, to their separation.
This dynamics may be written as $dZ/dt=-\gamma(Z-Z_0)$  (Newtonian creep), where $Z-Z_0$ is the height of one point above the equilibrium ellipsoid $Z_0$, normal to the ellipsoid, and $\gamma$ is a relaxation factor inversely proportional to the viscosity of the body at the surface. The equilibrium surface $Z_0(\phi,\lambda)$ is a triaxial ellipsoid ($\phi,\lambda$ are the spherical coordinates on the surface of the ellipsoid). 
The integration of this differential equation allows us to obtain the dynamical figure of equilibrium $Z(\phi,\lambda)$ and to define the boundaries of the integrals giving the torque due to the tidal forces as well as the disturbing forces acting on the system.

\section{Tidal evolution}\label{Evol}
Tidal evolution affects both the orbital elements of the pair and the rotation of the two bodies in interaction. The basic planar equations of this process showing how each of the two bodies contributes to the variations of the semi-major axis and eccentricity of the system are (Folonier et al., 2018, Online suppl.; Ferraz-Mello, 2019):

\begin{equation}\label{T1}
	[\langle \dot{a}\rangle]_i = \frac{3k_{2i}nm_jR_i^5} {m_ia^4}
	\left((1-5e^2)\frac{\gamma_i\nu_i}{\gamma_i^2+\nu_i^2} 
	-\frac{3e^2}{4}\frac{\gamma_i n}{\gamma_i^2+n^2} \hspace{3cm} \right.
\end{equation} 
\begin{displaymath} \left. \hspace{2cm}
	+\frac{e^2}{8}\frac{\gamma_i(\nu_i+n)}{\gamma_i^2+(\nu_i+n)^2}   
	+\frac{147e^2}{8}\frac{\gamma_i(\nu_i-n)}{\gamma_i^2+(\nu_i-n)^2}
	\right) + \mathcal{O}(e^4). 
\end{displaymath} 
and
\begin{equation}\label{T2}
	[\langle \dot{e}\rangle]_i = -\frac{3k_{2i}nem_jR_i^5}{4m_ia^5}
	\left( \frac{\gamma_i\nu_i}{\gamma_i^2+\nu_i^2} 
	+ \frac{3}{2} \frac{\gamma_i n}{\gamma_i^2+n^2} \hspace{4cm} \right.
\end{equation}
\begin{displaymath} \left.  \hspace{2cm}
	+\frac{1}{4}\frac{\gamma_i(\nu_i+n)}{\gamma_i^2+(\nu_i+n)^2}   
	-\frac{49}{4}\frac{\gamma_(\nu_i-n)}{\gamma_i^2+(\nu_i-n)^2} \right) 
	+ \mathcal{O}(e^3).
\end{displaymath}
where the subscript $i$ refers to the body deformed by the tidal stress 
and $j$ refers to the body whose gravitational attraction is creating the stress (they can be the star and the planet or vice versa); 
$a, e, n$ are the semi-major axis, eccentricity and mean-motion of the system; $m, R, \gamma$ are the masses, equatorial radii and
relaxation factors of the considered bodies, and
$$k_{2i}=\frac{15 C_i}{4m_i R_i^2}$$
($C_i$ is the moment of inertia).  The quantities
$$\nu_i = 2\Omega_i-2n$$ 
are the semi-diurnal frequencies.  
The rotation velocities of the bodies, $\Omega_i$ are affected by the tidal evolution and ruled by the equations
\begin{equation}\label{T3}
	\langle \dot{\Omega}_i\rangle = -\frac{3k_{2i}Gm_j^2R_i^5} {2C_i a^6}
	\left((1-5e^2)\frac{\gamma_i\nu_i}{\gamma_i^2+\nu_i^2} 
	+\frac{e^2}{4}\frac{\gamma_i(\nu_i+n)}{\gamma_i^2+(\nu_i+n)^2}   \right.
\end{equation} 
\begin{displaymath} \left. \hspace{2cm}
	+\frac{49e^2}{4}\frac{\gamma_i(\nu_i-n)}{\gamma_i^2+(\nu_i-n)^2}
	\right) + \mathcal{O}(e^4). 
\end{displaymath} 

These equations do not appear so simple as the corresponding equations found in the literature. The reason is that they are general. The parameters of extra-solar systems span various orders of magnitude and it is not possible to encompass all cases with  simplified formulas. If more accurate equations are needed, one may use Mignard's (1980) or Hut's (1981) versions of Darwin's CTL theory with rational functions of the eccentricities, instead of truncated series expansions  or, yet, the parametric version of the creep tide theory (Folonier et al. 2018; Ferraz-Mello et al. 2020) where the instantaneous polar oblateness, equatorial prolateness and lag are given by closed differential equations which may be integrated numerically together with the equations for the variation of the elements.

\subsection{Synchronisation}\label{synchro}
The right-hand side of eqn. (\ref{T3}) is dominated  by the term independent of the eccentricity. Given the negative sign of the coefficient in front of the brackets, the dominating term will have a sign contrary to the sign of the semi-diurnal frequency $\nu_i$. Then the tidal torque accelerates or brakes the rotation of the body so as to make it almost synchronous (exactly synchronous if $e=0$). The contribution of the terms dependent on the eccentricity, however, alters this rule. In order to see the actual location of the stationary rotation, let us simplify eqn. \ref{T3} by assuming that we may neglect terms of the order of $\nu_i/n$ and let us solve the equation $\langle \dot\Omega_i \rangle = 0$. There results
\begin{equation}\label{eq:st}
	\nu_{i \, (\rm stat)} = 12e^2\gamma_i\Psi_i + \mathcal{O}(e^4).
\end{equation}  
where 
\begin{equation}\label{eq:psi}
	\Psi_i=\frac{\gamma_i n}{\gamma_i^2+n^2}.
\end{equation}  
\begin{figure}[h!]
	\centerline{\hbox{
			\includegraphics[height=6cm,clip=]{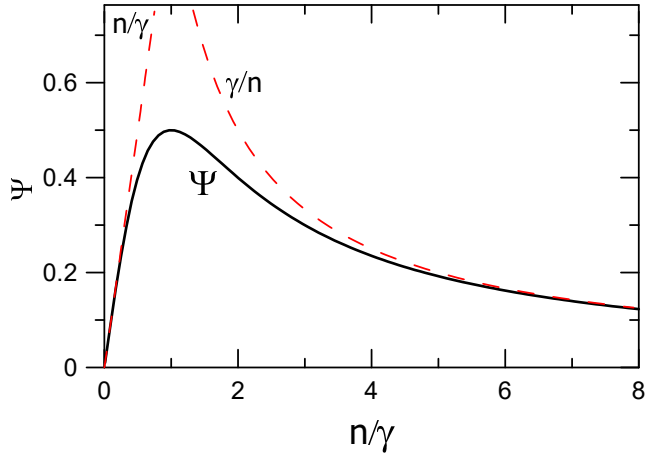}}}
	\caption{The function $\Psi=(\gamma/n + n/\gamma)^{-1}$ }  
	\label{fig:Psi}     
\end{figure}

Since $\Psi_i$ is positive and always less than 0.5, we see that the stationary value of $\nu_i$ is always positive. This means that $\Omega_{i\, (\rm stat}) > n$. Then, the final state of the rotation of one planet is not exactly synchronous, but supersynchronous, unless the eccentricity is equal to zero.

\subsection{Hot Jupiters and host stars}
More simple equations can be obtained in the case of systems harboring a close-in hot Jupiter (or other gaseous companions). In these cases, both the planet and the star have relaxation factors larger than 1 s$^{-1}$ (see section \ref{gammas}), such that, always, $\gamma_i \gg n$, allowing us to neglect terms of order $n/\gamma$. Hence, Eqns. (\ref{T1}-\ref{T2}) become
\begin{equation}\label{T4}
	[\langle \dot{a}\rangle]_i \simeq \frac{6k_{2i}n^2m_jR_i^5} {\gamma_i m_i a^4}
	\left((1+\frac{27}{2}e^2)\frac{\Omega_i}{n}- (1+23e^2)\right) + {\mathcal O}(e^4);
\end{equation}

\begin{equation}\label{T5}
	[\langle \dot{e}\rangle]_i = -\frac{27k_{2i}n^2em_jR_i^5}{\gamma_i m_ia^5}
	\left( 1-\frac{11}{18}\frac{\Omega_i}{n}\right) + {\mathcal O}(e^3).
\end{equation}

These equations are the same obtained with Darwin's theory when a constant time lag $\tau_i$ is assumed (see Hut, 1981; Ferraz-Mello et al. 2008), and the correspondence formula
\begin{equation}
	\tau_i=\frac{1}{\gamma_i}    \label{eq:tau}
\end{equation}
(valid when $\gamma_i \ll n$) is adopted. 

In this case, the synchronization process leads to a supersynchronous stationary rotation of the body whose angular velocity is the same found in Darwin's theory with the CTL hypothesis:
\begin{equation}
	\Omega_i=n+6ne^2.
\end{equation}

\subsection{Telluric exoplanets}
In the case of exoplanets with a surface composed at least partially of solid silicate material, like the Earth's crust, $\gamma$ is much smaller than in the case of gaseous planets. In the Solar System, the relaxation factor of terrestrial planets and ice satellites is $\gamma \ll 10^{-6} {\rm s}^{-1}$ (see Ferraz-Mello, 2019, Table 4). 

In addition, as the
synchronization of close-in telluric planets occurs quickly, we may introduce the additional hypothesis that the system reached the stationary state and fix a value for $\nu_i$ in accordance with eqn. (\ref{eq:st}). In this case, eqns. (\ref{T1}) and (\ref{T2}) become
\begin{equation}\label{T7}
	[\langle \dot{a}\rangle]_i \simeq -\frac{21k_{2i}ne^2m_jR_i^5} { m_i a^4}\ \Psi_i
\end{equation}

\begin{equation}\label{T8}
	[\langle \dot{e}\rangle]_i \simeq -\frac{21k_{2i}nem_jR_i^5}{2 m_i a^5}\ \Psi_i.
\end{equation}

It is worth noting that these equations are also valid for hot Jupiters trapped in stationary rotation, in which case, $\Psi_i \simeq n/\gamma_i$.


\section{Inclination and obliquity}
In the previous section, we collected the main perturbations in the orbital motion and in the planetary rotation, in the coplanar case, that is, when the rotation axis of the body is perpendicular to the orbital plane of the system. In the inclined case, one correction to the given formulas must be added. According to Darwin's CTL theory, to the formulas given above, we have to add (Hut, 1981; Ferraz-Mello et al. 2008) 

\begin{equation}\label{T9}
	[\langle \delta \dot{a} \rangle]_i \simeq -\frac{3k_{2i}nm_jR_i^5 \Omega_i} {\gamma_i m_i a^4}
	\sin^2 I_i,
\end{equation}

\begin{equation}\label{T10}
	\langle \delta\dot\Omega_i \rangle \simeq \frac{3k_{2i}Gm_j^2R_i^5 (\Omega_i-n)} {2C_i \gamma_i a^6}\sin^2 I_i.
\end{equation} 
where $I_i$ is the inclination of the equator of the tidally deformed body ($i$) with respect to the orbital plane. When this body is one planet moving around one star, the $I_i$ is often called {obliquity}. 
At the order considered, $[\langle \dot{\delta} e\rangle]_i \simeq 0$.

We note that the parameter $\tau_i$ of the CTL theories has been substituted by the inverse
of the relaxation factor $\gamma_i$ following what was established in Eqn. \ref{eq:tau}. It is also worth stressing the fact that CTL theories are only applicable to gaseous bodies (in which case, $\gamma_i \ll n$). 

\subsection{Tidal evolution of obliquity}
In the coplanar case, if we neglect the viscosity, the force is radial, its torque is equal to zero and, therefore, the rotation of the body is not affected by the static tide. There is no exchange of angular momentum between the rotation of the body and the orbital motion. Besides, the work done by the static tide force is an
exact differential and, therefore, the total mechanical energy of the system
remains constant in one cycle. There is no dissipation of energy (see discussion in Ferraz-Mello, 2019). The perturbations given in the previous section vanish if the viscosity of the body is neglected (that is, if we are in the limit $\gamma \rightarrow \infty$). 

In the spatial case, however, even in the static (or inviscid) case, the vertex of the ellipsoid representing the equilibrium figure is not directed to the companion (Folonier et al., 2022) and the torque no longer vanishes. However, the contributions of the tidal and of the rotational oblateness
may be considered separately and, assuming that the orbital plane has a constant precession (forced by a third body in an external orbit, for instance), the Cassini stationary states known in the case of rigid bodies may subsist. In this case, initially,  
if the star rotation is slower than the orbital motion, the tides raised by the planet on the star act decreasing the orbital semi-major axis. In the fall towards the planet, if  the planet is sufficiently far from the star, it may cross the situation that corresponds to the so-called Cassini 2 stationary state in which the fall of the planet towards the star brings as consequence an increase in the obliquity, which, in its turn, accelerates the fall to a point in which the permanence in the stationary state is no longer possible. The whole problem is very complex as it involves the variations of the equatorial and orbital planes due to tides in both bodies, and also external factors forcing an important precession of the orbital plane. A complete theory is not available. However, it may justify the existence of high-obliquity short-period planets (Fabricky et al. 2007;  Millholland et al. 2020.).

\section{Dissipation}
If one planet evolves according to the equations of section \ref{Evol}, the tides raised on the planet by the star change the orbital energy of the system and the rotational energy of the planet with the power
\begin{equation}
	[\langle \dot{E}_{\rm orb}\rangle]_p = \frac{3k_{2p}Gm_s^2nR_p^5} {2a^6}
	\left((1-5e^2)\frac{\gamma_p\nu_p}{\gamma_p^2+\nu_p^2} 
	-\frac{3e^2}{4}\frac{\gamma_p n}{\gamma_p^2+n^2} \right. \hspace{1cm}
\end{equation} 
\begin{displaymath} \left. \hspace{2cm}
	+\frac{e^2}{8}\frac{\gamma_p(\nu_p+n)}{\gamma_p^2+(\nu_p+n)^2}   
	+\frac{147e^2}{8}\frac{\gamma_p(\nu_p-n)}{\gamma_p^2+(\nu_p-n)^2}
	\right) + \mathcal{O}(e^4). 
\end{displaymath} 
and 
\begin{equation}
	[\langle \dot{E}_{\rm rot}\rangle]_p = -\frac{3k_{2p}Gm_s^2\Omega R_p^5} {2a^6}
	\left((1-5e^2)\frac{\gamma_p\nu_p}{\gamma_p^2+\nu_p^2} 
	+\frac{e^2}{4}\frac{\gamma_p(\nu_p+n)}{\gamma_p^2+(\nu_p+n)^2}  \right. 
\end{equation} 
\begin{displaymath} \left. \hspace{2cm}
	+\frac{49e^2}{4}\frac{\gamma_p(\nu_p-n)}{\gamma_p^2+(\nu_p-n)^2}
	\right) + \mathcal{O}(e^4). 
\end{displaymath} 
where the subscripts $p,s$ refer to the planet and the star, respectively (Ferraz-Mello, 2015).
These are the two only sources of the mechanical energy dissipated inside the planet. Strictly speaking, to have a complete picture, we should also consider the internal mechanical energy of the planet which changes in the process as the tidal forces affect the semi-axes of the ellipsoid representing the planet. However, this is a small quantity oscillating about zero. 
It is important to say that in the parametric version of the creep tide theory (Folonier et al. 2018), the mechanical energy variation appears as a sum of squares multiplied by a negative coefficient so that, in all circumstances,  
$ [\dot{E}_{\rm tot}]_p \le 0$. 

\begin{figure}[h]
	\centerline{\hbox{
			\includegraphics[height=5cm,clip=]{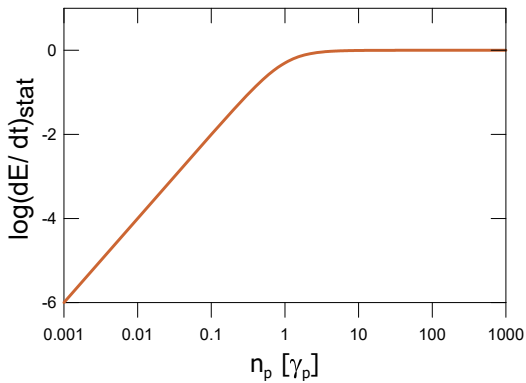}}}
	\caption{Power of the tidal energy dissipated in a planet in stationary rotation (arbitrary units) as a function of the orbital frequency (in units of $\gamma_p$). In the case of gaseous bodies, $n_p[\gamma_p] < 1 $ and, in the case of stiff planets, $n_p[\gamma_p] > 1 $.} 
	\label{fig:dEdt}
\end{figure}

Since the rotation of short-period planets is damped to the stationary state considered in section \ref{synchro} in a few million years, it is convenient to substitute the general formula with a much more simple one:
\begin{equation}\label{Estat}
	[\langle \dot{E}_{\rm stat}\rangle]_p = -\frac{21k_{2p}Gm_s^2n R_p^5} {2a^6}
	{e^2}\frac{\gamma_pn}{\gamma_p^2+n^2}
	+ \mathcal{O}(e^4)
\end{equation} 
(Folonier et al. 2018). This equation means that when $n \ll \gamma_p$ (e.g. the hot Jupiters), the  dissipation power grows with the square of the orbital frequency (i.e. the square of the mean motion) and when $n \gg \gamma$ (e.g. Earths and super-Earths),  the dissipation power does not depend on the mean motion (see fig. \ref{fig:dEdt}). This figure differs of the often shown $\Lambda$-shaped plot of the inverse of the quality factor $Q$. The reason is that in the latter case, $Q^{-1}$ is proportional to the energy dissipated in one period, and to get the average dissipation per unit time, we have to divide by the period.

In the inclined case, one correction to the given formulas must be added. According with Darwin's CTL theory, valid for gaseous bodies, we have to add, to the formulas given above, the term  

\begin{equation}
	[\langle \delta \dot{E}_{\rm stat}\rangle]_p = -\frac{3k_{2p}Gm_s^2n R_p^5} {2a^6}
	{\sin^2 I_p}\frac{n}{\gamma_p}
	+ \mathcal{O}(e^4)
\end{equation} 
(Ferraz-Mello et al. 2008) where $I_p$ is the inclination of the equator of the planet with respect to the orbital plane  ({obliquity}).

\section{Star wind braking}
\begin{figure}
	
	\centering
	
	\includegraphics[width=6cm]{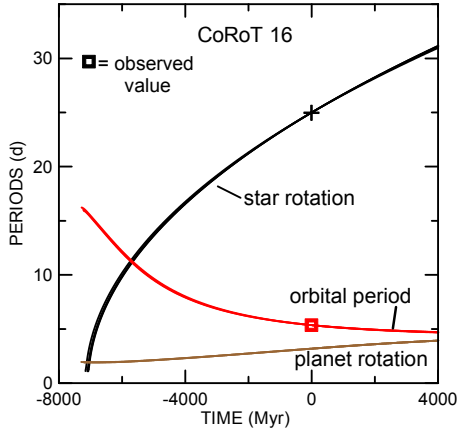}
	
	\caption {CoRoT-16: Long-term evolution of the stellar (black) and planetary (brown) rotational periods and of the orbital period of the planet (red).
		CoRoT-16 b is a hot Saturn in an orbit beyond the limit where the tidal interactions with the star influence significantly the stellar rotation. The known isochronal age of the star ($6.7 \pm 2.8$ Gyr) constrains the current stellar rotational period to the neighborhood of 25 days. 
		Taken from Ferraz-Mello (2016).} 
	\label{fig:C16}
	
\end{figure}

The rotational period of the host star plays a determinant role in the tidal orbital evolution of low-eccentricity  planets. The leading terms in Eqns. (\ref{T1}) and (\ref{T3}) are proportional to the semi-diurnal frequency $\nu=2\Omega-2n$ and thus have different signs when the orbital period is larger or smaller than the revolution period of the star\footnote
{This fact is at the origin of some incorrect generalizations. One quadratic term in the orbital eccentricity has a large coefficient which is negative if $|\nu_p| < n$. In addition, tides in the almost synchronous  planet give also a negative contribution to $\dot{a}$ (see Eqn. \ref{T7}). Hence, except for very low eccentricities, the planet will be falling onto the star no matter if the star rotation period is larger or smaller than the orbital period of the planet (see fig. \ref{fig:C16}).}. 
It happens that solar-type stars are prone to lose rotational angular momentum through the stellar wind. Initial periods of $1-2$ days in the early phases of the life of an active star may change to tens of days in a few Gyrs. Thus, one close-in planet in a low-eccentricity orbit that, at the beginning of its life was being ejected from the neighborhood of the star because of the tides raised by it on the star, after some time stops receding from the star and starts falling into it.
Therefore, the consideration of the star's angular momentum leakage is mandatory in long-term evolutionary studies. 

In order to consider the variations of the host star rotation, we may use Bouvier's model for the magnetic  braking of low-mass stars
($0.5{\rm M}_\odot < m_s < 1.1 {\rm M}_\odot$) with an outer convective zone: 
\begin{equation}
	\dot\Omega_s =	\left\{
	\begin{array}{ll}
		-  B_W\Omega_s^3&{\rm when}\qquad \Omega_s\le\omega_{\rm sat}\vspace{3mm}\\
		-  B_W\omega_{\rm sat}^2\Omega_s\hspace{1cm}& {\rm when}\qquad \Omega_s>\omega_{\rm sat}\label{eq:Bouvier}
	\end{array}
	\right.
	\ \end{equation}
where $\omega_{\rm sat}$ is the value at which the angular momentum loss saturates
(fixed at $\omega_{\rm sat}  = 3, 8, 14 \Omega_\odot$ for 0.5, 0.8, and 1.0 $M_\odot$ stars, respectively) and $B_W$ is a factor depending on the star moment of inertia ($C_s$), mass ($M_s$) and radius ($R_s$) through the relation
\begin{equation}
	B_W=2.7\times 10^{47} \frac{1}{C_s}\sqrt{\Big(\frac{R_s}{R_\odot} \frac{M_\odot}{M_s} \Big)}\qquad \qquad   ({\rm cgs \enspace units}).
	\label{eq:Bw}
\end{equation}
(Bouvier et al. 1997).
In the case of the Sun, the coefficient used in Eqn. (\ref{eq:Bouvier}) corresponds to $6.6  \times 10^{30}$ g cm$^2$ s$^{-2}$. 

For fully convective low-mass M-stars, more complex models are used in which the coefficient in Eqn. (\ref{eq:Bw}) is not the same for slow and fast rotators. Typical values in the case of slow rotators range from the same value as for solar-like stars to $1.1 \times 10^{47}$ \mbox{g cm$^2$ s,} depending on the adopted saturation value. For fast  rotators, the coefficient value is taken at least one order of magnitude smaller (see Irwin et al. 2011). 

The above form of the law is valid after the star has completed its contraction (the stellar moment of inertia $C_s$ no longer changes significantly) and is fully decoupled from its primeval disk. No significant mass loss needs to be considered. 

We may note that the magnetic connection to very close planets may inhibit the stellar wind (Strugarek et al. 2014) and thus reduce the braking torque on the star to a minimum of 70 percent of its normal value for  $a\simeq 3.5R_s$ (see op.cit. fig. 10) and request a customized model. In the study of the planet around the sub-giant star CoRoT-21, P\"atzold et al. (2012) have adopted a coefficient with a value smaller than the one adopted for solar-like stars.

F stars with masses larger than 1.3 $M_\odot$ are not expected to be affected by a magnetic braking.

\section{The relaxation factor}\label{gammas}
The parameters $\gamma_i$ are not ad hoc quantities, but they are not known. In order to allow the application of the given formulas to actual problems, we have to decide which values of $\gamma_i$ to use. 
The factors are discussed below for the several classes of objects considered.

\subsection{Stars}
Dissipation values of solar-type stars have been estimated by  Hansen (2010, 2012) from the analysis of the survival of short-period exoplanets. His results can be converted into the relaxation factor of the creep tide theory (Ferraz-Mello, 2013) through 
\begin{displaymath}
\gamma_s \simeq \frac{2 G k_{2s}|\nu_{s}|}{ 3n R_{s}^5 \sigma_{s}}
\end{displaymath}

His mean result $\sigma_s= 8.3 \times 10^{-64}$ g$^{-1}$ cm$^{-2}$ s$^{-1}$, for a wide range of stars, corresponds to  $\gamma_s \sim 5 - 45 {\ \rm s}^{-1}$ (assuming $|\nu_s/n| \sim 2$). This variation in $\gamma_s$ is mainly due to the strong  dependence on the radius and the larger values correspond to stars having half of the radius of the Sun. 

However, the study of transiting hot Jupiters around old stars in significantly non-circular orbits shows that $\gamma_s < 30 {\ \rm s}^{-1}$ often implies a too large exchange of angular momentum between the orbit and the star rotation inconsistent with  eccentricities not damped to zero.

Results in a range slightly higher, $\gamma_s \sim 20 - 80 {\ \rm s}^{-1}$, were derived from the comparison of the observed rotational and orbital periods in systems with big close-in companions and known age and rotation (Ferraz-Mello et al., 2015). In that cases, the rotational period of the star evolves under the action of the tides raised by the (big) close-in companion and, in the case of active stars, the wind braking. Their current value is critically determined by the relaxation factor of the star.   

\begin{figure}
	
	\centering
	
	\includegraphics[width=6cm]{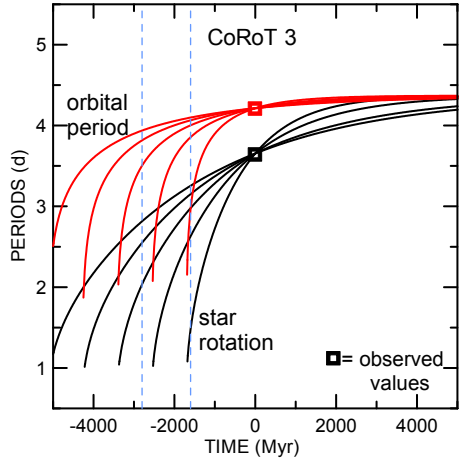}
	
	\caption {Simulations of the evolution of the rotational period of the host star CoRoT-3 (black) and of the orbital period of its companion, the brown dwarf  CoRoT-3 b (red). Adopted stellar tidal relaxation factors (starting from the steepest curves): 40, 60, 80, 100, 120 s$^{-1}$. The dashed vertical lines show the stellar age range. Taken from Ferraz-Mello (2016).}
	\label{fig:C3}
	
\end{figure}

One example is given in figure \ref{fig:C3} that shows results from simulations of the orbital evolution of the brown dwarf CoRoT-3 b and its host star CoRoT-3, a F3 star. No braking was added. In that example, in order to reproduce the known parameters of the system, we may have $40 < \gamma_s < 80 {\ \rm s}^{-1}$. A smaller (resp. larger) $\gamma_s$ means a faster (resp. slower) evolution and a system younger (resp. older) than given by the age of the star as determined from stellar models. 

The limit $\gamma_s > 30 {\ \rm s}^{-1}$ is also more or less the same obtained by Jackson et al. (2011) from the analysis of the distribution of the putative remaining lifetime of hot Jupiters. 

A lesser value is obtained from the decay of the planet WASP-12b. In that case, it was possible to fit a tidal model to the decreasing orbital period of the planet ($ 29 \pm  2$ msec/year) (Yee et al. 2019). The result obtained for the quality factor of the star corresponds to $\gamma_s = 18 {\ \rm s}^{-1}$.

\subsection{Hot Jupiters}
In his analyses, Hansen (2010) also made  
a first assessment of the dissipation of hot Jupiters. 
The comparison of his model to the creep tide theory, in this case, assumes that the planets are trapped in  stationary quasi-synchronous rotation. The correspondence formula is  
\begin{displaymath}
	\gamma_p \simeq \frac{3 G {k}_{2p} }{4  R_p^5 \sigma_p}
\end{displaymath}
where $\sigma_p$ is Hansen's planetary dissipation parameter. 
Then, using his mean results for WASP-17 b, CoRoT-5 b and Kepler-6 b, transiting planets whose radii have been determined, we obtain values of $\gamma_p$ in the range $10 - 60 {\ \rm s}^{-1}$, the smallest value corresponding to the bloated WASP-17 b  of radius $\sim 2R_{\rm Jup}$. 

In the case of CoRoT-5 b, a memory of the past evolution is kept by the eccentricity.
Backward simulations of this system show a significant rate of circularization and the eccentricity
must have been much larger in the past. If the relaxation factor is smaller than a certain limit, 
the resulting circularization rate would impose eccentricities close to 1 in the past Gyr (Ferraz-Mello, 2016), difficult to explain with our current knowledge of the exoplanets orbital evolution .

\subsection{Small planets}

We may transpose to exoplanets the results known for some small planets of our Solar System. We may mention the estimates given for Mercury,  $4-30 \times 10^{-9} {\ \rm s}^{-1}$ (Ferraz-Mello, 2015), and for the solid Earth,  $1-4 \times 10^{-7} {\ \rm s}^{-1}$ (conversion of the quality factor adopted by several authors).
We may add that the time lag necessary to a CTL theory to reproduce the tidal acceleration of the Moon is $\sim$600 s (Mignard, 1979), which is equivalent\footnote{For rocky planets, we cannot use the same correspondence formula $\gamma=1/\tau$ used for gaseous planets. In this case, we have to use the complete equivalence formula for $Q \simeq 1/\nu\tau$ where the function $\Psi$ is taken without approximation (Ferraz-Mello, 2013). Hence, the determination of $\gamma$ involves a second-degree equation with two solutions. For the whole Earth, besides the value given in the text, we also have the solution $\gamma \sim 1.6 \times 10^{-3} {\ \rm s}^{-1}$, the choice done being just an educated guess.} to $\gamma \sim 1.2 \times 10^{-5} {\ \rm s}^{-1}$.

However, it is not wise just to extend the relaxation factors estimated in the Solar System to most of the known exoplanets. The distances of telluric exoplanets to their host stars is generally too small and the stellar radiation power creates an extreme physical environment for the planet. We cannot exclude values different from the above-indicated ones. Nevertheless, the estimated viscosity of the super-Earth CoRoT-7 b is $\eta > 10^{18}$ \mbox{Pa s} (L\'eger et al., 2011), which corresponds, for this planet, to $\gamma_p < 5 \times 10^{-7} {\ \rm s}^{-1}$.   

We may compare these values to the analogous values for Neptune: $ 3-19 {\ \rm s}^{-1}$. The large difference in the order of magnitude of the relaxation factors of gaseous and stiff planets may be used to decide,
from the study of their putative tidal evolutions, if some potentially habitable small planet is  a super-Earth or a mini-Neptune of the same size (Gomes and Ferraz-Mello, 2020).

\section{Conclusion}

We visited the main formulas used in the study of the tidal evolution of one system formed by one star and one close-in planet taking into account the tides raised mutually in the two bodies. Most of these formulas are known, in one way or another, since the work of Darwin. The main ones shown here correspond to the approximation to ${\mathcal O} (e^2)$ of the hydrodynamical approach adopted by Ferraz-Mello (2012, 2013), the creep tide theory. Additionally, are also given, in the inclined problem, the formulas of Darwin's CTL-theory at the same order of approximation. 

The state of the tidal theories, in what concerns the inclination, is far from complete.
Classical theories consider only the terms proportional to the phase lag. As far as the coplanar case is considered, this is justified by the fact that the static tide (the deformation of the body considered as inviscid) is just a stretching of the body along the direction of the line joining the centers of the two bodies. Because of the symmetry of this configuration, the resulting torque is zero. In the tri-dimensional case with inclined rotation axes, however, this is not so. The bulge of the static tide in each body is no longer directed to its companion. It is pointed to a direction in the plane formed by the rotation axis of the body and the line joining the centers of the two bodies. The symmetry of the planar case is destroyed and the torque due to the static tide is no longer equal to zero (Folonier et al., 2022). Therefore, the contribution of the static tide cannot be neglected. In fact, it should not be neglected even in the coplanar case where it does not give long-term contributions to the energy and angular momentum, but affects the angular elements of the orbit which will precess because of the static tide.

We also added a summary of the numerical values of the relaxation factors. Some early estimations of the dissipation in stars and hot Jupiters (Hansen, 2010) were used to determine the relaxation factors of those objects. The relaxation factor of the star WASP-12 could also be known from the dissipation determined from the observed decay of the orbit of its planet. In the case of stars with large companions (hot Jupiters or brown dwarfs) and with well-determined physical parameters (mass, radius, age, spin), the tides in the host star affect its period of rotation and models using the given equations allow an estimation of the $\gamma_s$ (Ferraz-Mello et al., 2015). We have, now, a large number of bright and well-known host stars with large companions and we can expect that the reported investigations be soon replicated with larger samples thus improving our knowledge of the tidal parameters of stars. A new analysis allowing a better knowledge  of the tidal parameters of the planets would also be welcome. 

\begin{acknowledgement}
	To Maurice Ravel, ever-present companion during this endless quarantine. I thank Prof. C. Beaug\'e, Dr. H. A. Folonier and G. O. Gomes for all discussions and suggestions. This investigation is sponsored by 
	CNPq (Proc. 303540/2020-6) and FAPESP (Proc. 2016/13750-6 ref. PLATO mission).
	
\end{acknowledgement}


\begin{thebibliography}{}

\bibitem[]{Bou}
{Bouvier, J., Forestini, M., Allain, S.}  1997, 
``The angular momentum evolution of low-mass stars". 
\textit{A\&A} {326}, 1023-1043

\bibitem[]{D79} 
{Darwin, G.H.} 1879, 
``On the bodily tides of viscous and semi-elastic spheroids and on the ocean tides upon a yielding nucleus''. 
\textit{Philos. Trans.} {170}, 1-35 (repr. Scientific Papers, Cambridge, Vol. II, 1908)

\bibitem[]{D80} 
{Darwin, G.H.} 1880,
``On the secular change in the elements of the orbit of a satellite revolving about a tidally distorted planet''.
\textit{Philos. Trans.} {171}, 713-891 (repr. Scientific Papers, Cambridge, Vol. II, 1908)

\bibitem[]{Efr} 
{Efroimsky, M., Lainey, V.} 2007, 
``Physics of Bodily Tides in Terrestrial Planets and the Appropriate Scales of Dynamical Evolution".
\textit{J. Geophys. Res.}{112}, E12003

\bibitem[]{Fabr}
{Fabrycky, D.C., Johnson, E.T., Goodman, J.} 2007,
``Cassini states with dissipation: Why obliquity tides cannot inflate hot Jupiters". 
\textit{ApJ} 665, 754.

\bibitem[]{FRH}
{Ferraz-Mello, S., Rodr\'{\i}guez, A., Hussmann, H.} 2008,
``Tidal friction in close-in satellites and exoplanets. The Darwin theory re-visited".
\textit{Cel.Mech.Dyn.Astr.} {101}, 171-201. Errata: \textit{Cel.Mech.Dyn.Astr.} {104}, 319-320 (2009).
(ArXiv: 0712.1156) 

\bibitem[]{RhDDA}
{Ferraz-Mello, S.} 2012,
``Dissipation and Synchronization due to creeping tides". 
\textit{AAS/DDA meeting} 43, id.8.06 (ArXiv: 1204.3957)

\bibitem[]{Rh1}
Ferraz-Mello, S., 2013,
``Tidal synchronization of close-in satellites and exoplanets. A rheophysical approach". 
\textit{Cel.Mech.Dyn.Astr.} {116}, 109-140. 

\bibitem[]{Rh2}
{Ferraz-Mello, S.} 2015,  
``Tidal synchronization of close-in satellites and exoplanets: II. Spin dynamics and extension to Mercury and exoplanets host stars".
\textit{Cel.Mech.Dyn.Astr.} {122}, 359-389. Errata: \textit{Cel.Mech.Dyn.Astr.} {130}:78 (2018), pp. 20-21. (arXiv: 1505.05384)  

\bibitem[]{host}
{Ferraz-Mello, S., Folonier, H., Tadeu dos Santos, M., Csizmadia, Sz., do Nascimento, J.D., P\"atzold, M.} 2015,  
``Interplay of tidal evolution and stellar wind braking in the rotation of stars hosting massive close-in planets". 
\textit{ApJ} {807}, 78 

\bibitem[]{Lega}
{Ferraz-Mello, S.} 2016,
``Tidal evolution of CoRoT massive planets and brown dwarfs and of their host stars". 
In A. Baglin and CoRoT team (eds.) \textit{The CoRoT Legacy Book}, EDP Sciences,  pp. 169-176. 

\bibitem[]{LecT}
{Ferraz-Mello, S.} 2019, 
``Planetary tides: theories". 
In G.Ba\`u et al. (eds.) \textit{Satellite Dynamics and Space Missions}, Springer, pp. 1-50. 

\bibitem[]{EPJ}
{Ferraz-Mello, S., Beaug\'e, C., Folonier, H.A., Gomes, G.O.} 2020,
``Tidal friction in satellites and planets. The new version of the creep tide theory". 
\textit{Eur. Phys. J. ST}, 229, 1441-62.

\bibitem[]{Rh3}
{Folonier, H.A., Ferraz-Mello, S., Andrade-Ines, E.} 2018,  
``Tidal synchronization of close-in satellites and exoplanets: III. Tidal dissipation revisited and application to Enceladus"
\textit{Cel. Mech. Dyn. Astr.} 130: 78 (arXiv: 1707.09229v2)

\bibitem[]{Fol3D}
{Folonier, H.A., Bou\'e, G., Ferraz-Mello, S.} 2022,
``Ellipsoidal equilibrium figure and Cassini states of rotating planets and satellites deformed by a tidal potential in the spatial case".
\textit{Cel.Mech.Dyn.Astr.} 134: 1

\bibitem[]{Yeva}
{Gevorgyan, Y., Bou\'e, G., Ragazzo, C., Ruiz, L.S., Correia, A.C.} 2020, 
``Andrade rheology in time-domain. Application to Enceladus' dissipation of energy due to forced libration". 
\textit{Icarus}, 343, p. 113610.

\bibitem[]{Gomes}
{Gomes, G.O., Ferraz-Mello, S.} 2020,
``Tidal evolution of exoplanetary systems hosting potentially habitable exoplanets. The cases of LHS-1140 bc and K2-18 bc". 
\textit{MNRAS} 494, 5082-5090

\bibitem[]{Hans10}
{Hansen, B.M.S.} 2010,
``Calibration of equilibrium tide theory for extrasolar planetary systems".
\textit{ApJ} {723}, 285-299

\bibitem[]{Hans12}
{Hansen, B.M.S.} 2012,
``Calibration of equilibrium tide theory for extrasolar planetary systems. II".
\textit{ApJ} {757}: 6

\bibitem[]{Hut}
{Hut, P.} 1981,
``Tidal evolution in close binary systems". 
\textit{A\&A}  99, 126-140

\bibitem[]{Irw}
{Irwin, J., Berta, Z. K., Burke, C., Charbonneau, D., Nutzman, P. et al.}  2011,
``On the angular momentum evolution of fully convective stars: rotation periods for field M-dwarfs from the MEarth transit survey".
\textit{ApJ} {727}: 56

\bibitem[]{JacGB}
{Jackson, B., Greenberg, R., Barnes, R.} 2008,
``Tidal evolution of close-in extrasolar planets. 
\textit{ApJ} 678, 1396.

\bibitem[]{JacPB}
{Jackson, B., Penev, K., Barnes, R.} 2011,
``Constraining Tidal Dissipation in Stars and Destruction Rates of Exoplanets",
\textit{BAAS} {43}, \#402.06

\bibitem[]{Kau}
{Kaula, W.M.} 1964, 
``Tidal dissipation by solid friction and the resulting orbital evolution".
\textit{Rev. Geophys.} {3} 661-685

\bibitem[]{Leg}
{L\'eger, A., Grasset, O., Fegley, B., Codron, F., Albarede, A.F. et al.} 2011,
``The extreme physical properties of the CoRoT-7b super-Earth". 
\textit{Icarus} 213, 1-11

\bibitem[]{Mac} 
{MacDonald, G.F.} 1964, 
``Tidal Friction", 
\textit{Rev. Geophys.} {2}, 467-541

\bibitem[]{Mig1} 
{Mignard, F.} 1979, 
``The evolution of the lunar orbit revisited - I". 
\textit{Moon and Planets}  {20}, 301-315

\bibitem[]{Mig2} 
{Mignard, F.} 1980, 
``The evolution of the lunar orbit revisited - II". 
\textit{Moon and Planets} {23}, 185-201

\bibitem[]{Mill20}
{Millholland, S.C., Spalding, C.} 2020,  
``Formation of ultra-short-period Planets by Obliquity-driven Tidal Runaway". 
\textit{ApJ} 905: 71

\bibitem[]{Paet}
{P\"atzold, M., Endl, M., Csizmadia, Sz., Gandolfi, D., Jorda, L. et al.}  2012, 
``Transiting Exoplanets from the CoRoT Space Mission XXIII. CoRoT-21b: a doomed large Jupiter arount a faint subgiant star". 
\textit{A\&A}  545: A6.

\bibitem[]{Stru}
{Strugarek, A., Brun, A.S., Matt, S.P., R\'eville, V.}  2014,
``On the diversity of magnetic interactions in close-in star-planet systems".
\textit{ApJ} 795: 86

\bibitem[]{Yee}
{Yee, S.W., Winn, J.N., Knutson, H.A., Patra, K.C., Vissapragada, S. et al.} 2019, 
``The orbit of WASP-12b is decaying".
\textit{ApJ} (Letters) 888: L5.


\end{thebibliography}
\end{document}